\documentclass[aps,preprint]{revtex4}%
\usepackage{amsfonts}
\usepackage{amsmath}
\usepackage{amssymb}
\usepackage{graphicx}
\usepackage{pdfpages}
\usepackage{float}
\usepackage{color}
\usepackage{xcolor}
\usepackage{caption}
\usepackage{lineno,hyperref}
\usepackage{changes}%
\hypersetup{colorlinks =true, allcolors = blue}
\providecommand{\U}[1]{\protect\rule{.1in}{.1in}}

\begin{document}

\begin{center}
{\Large Probing regular MOG static spherically symmetric spacetime using greybody factors and quasinormal modes}

Ahmad Al-Badawi

Department of Physics, Al-Hussein Bin Talal University, P. O. Box: 20,
71111, Ma'an, Jordan

\bigskip E-mail: ahmadbadawi@ahu.edu.jo

{\Large Abstract}
\end{center}

We investigate the behavior of the regular modified gravity (MOG) static
spherically symmetric black hole (BH) under massless scalar perturbation,
gravitational perturbation, and massless Dirac perturbation. The
dimensionless parameter $\left( \alpha \right) $ distinguishes this BH from
a Schwarzschild BH. We derive the effective potential equations for
three perturbations in the regular MOG BH. Using the derived potentials, we
calculate the bounds of greybody factors (GFs). Next, we investigate the
quasinormal mode (QNM) of the MOG BH by implementing the WKB method of sixth
order. By analyzing the influence of the MOG parameter $\alpha $ for the
BH we study on GF and QNM, we found that as $\alpha $ increases, the GFs
increase proportionally. However, both
gravitational wave oscillation frequency and damping decrease as $\alpha $ increases. Moreover,
we examine the behavior of QNMs by considering how their frequency changes
with the shape of potentials. As a result, we found that the frequency
behavior is like the quantum mechanical one. The faster the wave
decays, the larger the potential.

\section{Introduction}

In the event that LIGO confirms the existence of gravitational waves, the field of black hole (BH) physics becomes immensely rich for researchers \cite{qn1,qn2,qn3}. It is possible to distinguish between alternatives to BHs, such as quantum effects induced dark compact objects, quantum gravity and BHs, by examining binary BHs, remnants of compact binary star systems, and neutron stars culminating in the ringdown phase \cite{qn4}. There may be
no horizons and essential singularities on those compact objects. In
General Relativity (GR), BHs are spaces in spacetime where classical physics
breaks down at the BH's essential singularity. Even though GR has proven to
be successful, it is not without faults. There are two major weaknesses in
this theory: the existence of singularities \cite{qn5,qn6} and the absence of
observations verifying the existence of dark matter \cite{qn7}. Regarding this
issue, there are two groups of researchers \cite{qn8,qn9,qn10,qn11,qn12}. It is either dark matter
exists, or the theory of GR must be modified. It has been reported that search for dark matter has been unsuccessful in all experimental attempts \cite{qn13,qn14}. We therefore need a theory that removes the above mentioned ambiguities
in order to explore the nature of BHs. Different approaches may be used to
restructure its geometrical part of GR, which can, for example, be used.
Moffat proposed and developed the Scalar-Tensor-Vector (STVG) theory that
describes gravitational interaction \cite{qn15}, the so-called Modified Gravity
(MOG). MOG theory can explain several astrophysical observations
\cite{qn16,qn17,qn18,qn19,qn20,qn21,qn22}. The theory is also successful at describing the structure growth, matter power
spectrum, and the acoustical and angular power spectrum
of the cosmic microwave background \cite{qn23,qn24}.

In a recent paper, Moffat deduced a generalized Kerr rotating MOG dark compact object based on angular momentum (spin), mass, and parameter $\alpha$ \cite{qn25}. When $\alpha >\alpha _{cirt}=0.674$, MOG
spinning compact object and static compact object can be regular without
horizons and with and without an ergosphere, depending on their spin
parameter $a$ and coordinate $\theta $. We focus on a regular MOG static
spherically symmetric BH. The dimensionless parameter $\alpha$ distinguishes this BH from a Schwarzschild BH. A recent study of the shadow behavior of MOG
dark compact objects is presented in \cite{qn26}. In this study, the shadow for the
regular MOG BH has been analyzed and compared with M87* and Sgr A* data. Further,
they notice that as the MOG parameter increases, the radius of the photon
sphere, the radii of the shadow and the event horizon decrease. Using the Lagrangian formalism, \cite{qn27} explores both electrically neutral and charged particle motion in the MOG BH spacetime. They
found that increasing the value of the MOG parameter increases the effective
potential of the neutral particle moving in the spacetime of the regular MOG
dark compact object, but not the effective potential of the electrically
charged particle far from the source.

The purpose of this study is to investigate the physical properties of a
regular static spherically symmetric MOG dark compact object \cite{qn25} using
greybody factors (GFs) and quasinormal modes (QNMs) under massless scalar
fields, axial gravitational, and massless Dirac perturbations. Physicist
Hawking showed in 1974 that BHs are not perfectly black, but emit
particles \cite{qn28,qn29} along with scattering, absorbing, and radiating. In the
presence of a BH, Hawking radiation propagates on curved spacetime.
Spacetime's curvature acts like a gravitational potential, scattering
radiation. Thus, part of radiation is reflected into the BH, and
part transmitted to spatial infinity. The transmission probability or GF can
be calculated by several different methods, including matching technique
\cite{qn30,qn31,qn32}, WKB approximation \cite{qn33,qn34,qn35}, finding Bogoliubov coefficients method
\cite{qn36,qn37,qn38,qn39}, Miller-Good transformation method \cite{qn40}, and rigorous bounds
\cite{qn41,qn42}.

The perturbation of a BH produces gravitational waves dominated by QNM \cite{qn43}. In the QNM frequency, the real part represents the oscillation frequency of the perturbed BH, while the imaginary part represents the decay rate\cite{qn44,qn45,qn46,qn47,qn48}. The QNM frequency is also a significant factor in determining
the parameters of BHs as well as determining their spacetime stability.
LIGO/VIRGO detected gravitational waves produced by merging BHs for the first time in 2016. Gravitational wave signals are characterized by oscillations
that rapidly decay at the end of the waveform, known as a ringdown \cite{qn48a,qn48b,qn48c}. It is the ringdown phase i.e., QNM of the remnant BH which makes BH QNM research attractive.

This work is motivated by the following idea: we can gain insight into BH behavior in MOG theory and differentiate
it from GR by probing regular MOG static
spherically symmetric spacetime with GFs and QNM. Furthermore,  because GFs and QNMs play a significant role in BH physics, their analyses can also provide valuable insights. Furthermore, 
the QNMs might be detectable by new generations of gravitational wave detectors, and so might provide some clues to constructing a MOG theory. The paper
is organized as follows. Sect. II briefly reviewed the regular MOG static
spherically symmetric spacetime. Furthermore, we analyze the scalar field,
axial gravitational and massless Dirac field perturbations. Sect. III is
devoted to the calculation of bounds of the GFs of the regular MOG BH and
analyzing their graphical behavior. In Sect. IV, the WKB method of sixth order is
used to analyze QNMs. The effect of MOG parameter on QNMs is
analyzed for MOG BH. Sect. V contains a summary of the main conclusions.

\section{Regular MOG BH and Perturbations}

\subsection{ A brief review of the spacetime}

The MOG regular, static spherically symmetric solution can be written as
\cite{qn25,qn49} 
\begin{equation}
ds^{2}=-f\left( r\right) dt^{2}+f^{-1}(r)dr^{2}+r^{2}\left( d\theta
^{2}+\sin ^{2}\theta d\phi ^{2}\right)   \label{M1}
\end{equation}%
where 
\begin{equation}
f\left( r\right) =1-\frac{2\left( 1+\alpha \right) Mr^{2}}{\left(
r^{2}+\alpha \left( 1+\alpha \right) M^{2}\right) ^{3/2}}+\frac{\alpha
\left( 1+\alpha \right) M^{2}r^{2}}{\left( r^{2}+\alpha \left( 1+\alpha
\right) M^{2}\right) ^{2}},
\end{equation}%
in which, $M$ is the mass parameter of the gravitating object and $\alpha $
is the MOG parameter. Setting $\alpha =0$, the metric (\ref{M1})
reduces to the Schwarzschild BH. Such a regular MOG static spherically
symmetric spacetime might have two, one or no event horizon(s) depending on
the parameter, $\alpha $ \cite{qn26}. Figure 1 shows the radial dependence of the metric
function $f(r)$\ for different values of the MOG parameter $\alpha $. As illustrated in the figure, metric (\ref{M1}) admits three types of BH: no horizon $%
\left( \alpha >\alpha _{cirt}=0.674\right) $ , one horizon $\left( \alpha
=\alpha _{cirt}\right) $ and two horizons $\left( \alpha <\alpha
_{cirt}\right) .$ The metric function $f(r)$\ behaves as%
\begin{equation}
\lim_{r\rightarrow 0}f(r)=1+\frac{r^{2}}{M^{2}}\left( \frac{\alpha -2\sqrt{%
\alpha \left( 1+\alpha \right) }}{\alpha ^{2}\left( 1+\alpha \right) }%
\right) +\mathcal{O}\left( r^{3}\right) .  \label{mf1}
\end{equation}%
\begin{equation}
\lim_{r\rightarrow \infty }f(r)=1-\frac{2\left( 1+\alpha \right) M}{r}+\frac{%
\alpha \left( 1+\alpha \right) M^{2}}{r^{2}}+\mathcal{O}\left( \frac{1}{r^{3}}\right) .
\label{mf2}
\end{equation}%
When approaching the source, Eq. (\ref{mf1}) reduces to $1$, indicating that
MOG BH is regular. However when $\alpha <\alpha _{cirt},$ Eq. (\ref{mf2})
has two horizons given by 
\begin{equation}
r_{\pm }=M\left( 1+\alpha \pm \sqrt{1+\alpha }\right) .
\end{equation}
\begin{figure}
\centering
\includegraphics[scale=1]{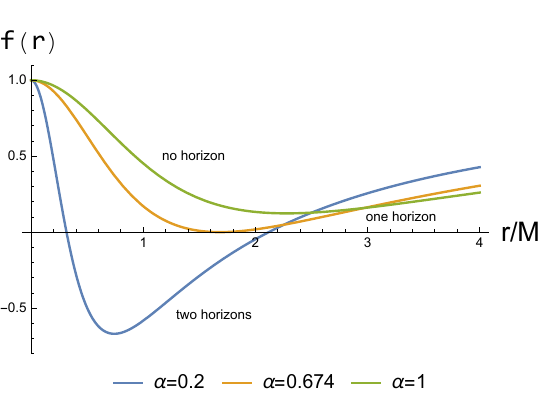} \caption{ Metric function of MOG BH for different values of the MOG parameter $\alpha$.}
\label{Figure1}%
\end{figure}

\subsection{Scalar Perturbation}

In this subsection, we study scalar perturbations in MOG BH backgrounds by
including test fields and solving the equations for specific test
fields. In curved spacetime, the massless scalar field is represented by the Klein-Gordon equation 
\begin{equation}
\frac{1}{\sqrt{-g}}\partial _{\mu }\sqrt{-g}g^{\mu \nu }\partial _{\nu
}U(t,r,\theta ,\phi )=0.  \label{is2}
\end{equation}%
For the spherical background each multipole
moment of the perturbation field evolves separately as
\begin{equation}
U(t,r,\theta ,\phi )=\frac{1}{r}Y_{m}^{l}\left( \theta ,\phi \right) \exp
\left( -i\omega t\right) ,
\end{equation}%
where $Y_{m}^{l}$ are the usual spherical harmonics. Putting our considered MOG BH metric into Eq. (\ref{is2}), we  get for each
multipole moment of the perturbation field the following equations
of motion. 
\begin{equation}
\left( \frac{d^{2}}{dr_{\ast }^{2}}+\frac{d^{2}}{dt^{2}}+V_{eff}\right) U=0,
\label{s1}
\end{equation}%
where $r_{\ast }$ is the tortoise coordinate: $\frac{dr_{\ast }}{dr}=\frac{1%
}{f},$ and $V_{eff}$ is the effective potential given by%
\begin{equation}
V_{eff}=\left( 1-\frac{2\left( 1+\alpha \right) Mr^{2}}{\left( r^{2}+\alpha
\left( 1+\alpha \right) M^{2}\right) ^{3/2}}+\frac{\alpha \left( 1+\alpha
\right) M^{2}r^{2}}{\left( r^{2}+\alpha \left( 1+\alpha \right) M^{2}\right)
^{2}}\right) \times  \label{v1}
\end{equation}%
\begin{equation*}
\left( \frac{l(l+1)}{r^{2}}+\frac{2\alpha M^{2}\left( 1+\alpha \right) }{%
\left( r^{2}+\alpha \left( 1+\alpha \right) M^{2}\right) ^{2}}-\frac{%
4M\left( 1+\alpha \right) }{\left( r^{2}+\alpha \left( 1+\alpha \right)
M^{2}\right) ^{3/2}}\right.
\end{equation*}%
\begin{equation*}
\left. +\frac{6Mr^{2}\left( 1+\alpha \right) }{\left( r^{2}+\alpha \left(
1+\alpha \right) M^{2}\right) ^{5/2}}-\frac{4\alpha M^{2}r^{2}\left(
1+\alpha \right) }{\left( r^{2}+\alpha \left( 1+\alpha \right) M^{2}\right)
^{3}}\right) .
\end{equation*}%
where $l$ is the angular quantum number.

\subsection{ Axial gravitational perturbations}

Gravitational perturbations are usually introduced into the background
metric $g_{\mu \nu }^{0}$ by introducing a small perturbation $h_{\mu \nu }$
as \cite{qn50}
\begin{equation}
g=g_{\mu \nu }^{0}+h_{\mu \nu },\left\vert h_{\mu \nu }\right\vert <<1.
\end{equation}%
Furthermore, the Ricci tensor become%
\begin{equation}
R_{\mu \nu }=R_{\mu \nu }^{0}+\delta R_{\mu \nu },\delta R_{\mu \nu }=\delta
\Gamma _{\mu \alpha ;\nu }^{\alpha }-\delta \Gamma _{\mu \nu ;\alpha
}^{\alpha },
\end{equation}%
where%
\begin{equation}
\delta \Gamma _{\mu \nu }^{\beta }=\frac{1}{2}g^{\beta \alpha }\left(
h_{\alpha \nu ;\mu }+h_{\alpha \mu ;\nu }-h_{\mu \nu ;\alpha }\right) ,
\end{equation}%
It is possible to separate the equations describing axial perturbations when
the metric perturbation tensor $h_{\mu \nu }$ is expanded into tensor
spherical harmonics\ \cite{qn51}
\begin{equation}
h_{\mu \nu }=\left( 
\begin{array}{cccc}
0 & 0 & 0 & h_{0} \\ 
0 & 0 & 0 & h_{1} \\ 
0 & 0 & 0 & 0 \\ 
h_{0} & h_{1} & 0 & 0%
\end{array}%
\right) \sin \theta \frac{\partial }{\partial \theta }P_{l}\left( \cos
\theta \right) ,
\end{equation}%
where $P_{l}\left( \cos \theta \right) $ is the Legendre polynomial and the
functions $h_{0}\left( t,r\right) $,$h_{1}\left( t,r\right) $ satisfy the
following equations%
\begin{equation}
\frac{1}{f}\frac{dh_{0}}{dt}-\frac{df}{dr}h_{1}-\frac{dh_{1}}{dr}f=0,
\end{equation}%
\begin{equation*}
\frac{d^{2}h_{1}}{dt^{2}}-\frac{d^{2}h_{0}}{dtdr}+\frac{2}{r}\frac{dh_{0}}{dt}+f\frac{%
l^{2}+l-2}{r^{2}}h_{1}=0,
\end{equation*}%
\begin{equation*}
\frac{d^{2}h_{0}}{dr^{2}}-\frac{d^{2}h_{1}}{dtdr}-\frac{2}{r}\frac{dh_{1}}{dt%
}+\frac{2}{r^{2}f}\left( r\frac{df}{dr}-\frac{1}{2}\left( l^{2}+l\right)
\right) h_{0}=0.
\end{equation*}%
Defining $U(t,r)=\frac{1}{r}fh_{1},$ we can obtain the Schrodinger-like
equation of the axial gravitational perturbation via the tortoise
coordinate, namely
\begin{equation}
\left( \frac{d^{2}}{dr_{\ast }^{2}}+\frac{d^{2}}{dt^{2}}+V_{eff}\right) U=0,
\label{s2}
\end{equation}%
with the potential given by 
\begin{equation}
V_{eff}=\left( 1-\frac{2\left( 1+\alpha \right) Mr^{2}}{\left( r^{2}+\alpha
\left( 1+\alpha \right) M^{2}\right) ^{3/2}}+\frac{\alpha \left( 1+\alpha
\right) M^{2}r^{2}}{\left( r^{2}+\alpha \left( 1+\alpha \right) M^{2}\right)
^{2}}\right) \times  \label{vg1}
\end{equation}%
\begin{equation*}
\left( \frac{l(l+1)}{r^{2}}-\frac{6\alpha M^{2}\left( 1+\alpha \right) }{%
\left( r^{2}+\alpha \left( 1+\alpha \right) M^{2}\right) ^{2}}+\frac{%
12M\left( 1+\alpha \right) }{\left( r^{2}+\alpha \left( 1+\alpha \right)
M^{2}\right) ^{3/2}}\right.
\end{equation*}%
\begin{equation*}
\left. -\frac{18Mr^{2}\left( 1+\alpha \right) }{\left( r^{2}+\alpha \left(
1+\alpha \right) M^{2}\right) ^{5/2}}+\frac{12\alpha M^{2}r^{2}\left(
1+\alpha \right) }{\left( r^{2}+\alpha \left( 1+\alpha \right) M^{2}\right)
^{3}}\right) .
\end{equation*}

\subsection{Dirac perturbations}

To study the massless Dirac fields propagating in regular MOG
static spherically symmetric space, we will use the Newman-Penrose formalism
\cite{qn52}. The Chandrasekhar-Dirac (CD) equations for massless Dirac field \cite{qn53} in the Newman-Penrose formalism are
given by%
\begin{equation*}
\left( D+\epsilon -\rho \right) F_{1}+\left( \overline{\delta }+\pi -\alpha
\right) F_{2}=0,
\end{equation*}%
\begin{equation*}
\left( \Delta +\mu -\gamma \right) F_{2}+\left( \delta +\beta -\tau \right)
F_{1}=0,
\end{equation*}%
\begin{equation*}
\left( D+\overline{\epsilon }-\overline{\rho }\right) G_{2}-\left( \delta +%
\overline{\pi }-\overline{\alpha }\right) G_{1}=0,
\end{equation*}%
\begin{equation}
\left( \Delta +\overline{\mu }-\overline{\gamma }\right) G_{1}-\left( 
\overline{\delta }+\overline{\beta }-\overline{\tau }\right) G_{2}=0,\label{cd1}
\end{equation}%
where $F_{1},F_{2},G_{1}$and $G_{2}$ represent the Dirac spinors, the letters $\rho ,\mu ,\epsilon ,\tau
,\gamma ,,\beta ,$ and $\alpha $ are the spin coefficients and
the bar denotes complex conjugation. Let us introduce the following basis vectors of null tetrad
in terms of elements of the metric (\ref{M1}) as 
\begin{equation*}
l^{\mu }=\left( \frac{1}{f},1,0,0\right) ,\qquad n^{\mu }=\frac{1}{2}\left(
1,-f,0,0\right) ,
\end{equation*}
\begin{equation}
m^{\mu }=\frac{1}{\sqrt{2}r}(0,0,1,\frac{i}{\sin \theta }),\qquad \overline{m%
}^{\mu }=\frac{1}{\sqrt{2}r}(0,0,1,\frac{-i}{\sin \theta }).  \label{7}
\end{equation}%
Hence, the directional derivatives in Eqs. (\ref{cd1})  are defined by $D=l^{\mu
}\partial _{\mu },\Delta =n^{\mu }\partial _{\mu }$ and $\delta =m^{\mu
}\partial _{\mu }$. The spin coefficients can then be computed as 
\begin{align}
\rho & =-\frac{1}{r},\mu =-\frac{f}{2r},\epsilon =\tau =0\qquad   \notag \\
\gamma & =\frac{f^{\prime }}{4},\beta =-\alpha =\frac{\cot \theta }{2\sqrt{2}%
r}.  \label{8}
\end{align}
Using equations (\ref{7}) and (\ref{8}) in CD equations (\ref{cd1}) leads to%
\begin{equation*}
\left( \emph{D}-\frac{1}{r}\right) F_{1}+\frac{1}{\sqrt{2}r}\mathit{L}%
F_{2}=0,
\end{equation*}%
\begin{equation*}
\frac{-f}{2}\left( \mathit{D}^{\dag }-\frac{f^{\prime }}{2f}+\frac{1}{r}%
\right) F_{2}+\frac{1}{\sqrt{2}r}\mathit{L}^{\dag }F_{1}=0,
\end{equation*}%
\begin{equation*}
\left( \emph{D}+\frac{1}{r}\right) G_{2}-\frac{1}{\sqrt{2}r}\mathit{L}^{\dag
}G_{1}=0,
\end{equation*}%
\begin{equation}
\frac{f}{2}\left( \mathit{D}^{\dag }-\frac{f^{\prime }}{2f}+\frac{1}{r}%
\right) G_{1}+\frac{1}{\sqrt{2}r}\mathit{L}G_{2}=0,  \label{D10}
\end{equation}%
where
\begin{equation}
\mathit{D}^{\dag }=-\frac{2}{f}\Delta,\mathit{L}=\sqrt{2}r\overline{\delta }+\frac{\cot \theta }{2},\mathit{L}^{\dag }=\sqrt{2}r\delta +\frac{\cot \theta }{2}.
\end{equation}%
To solve the CD equations (\ref{D10}), we consider the massless Dirac fields in the form 
\begin{equation}
F_{1}=R_{1}\left( r\right) A_{1}\left( \theta \right) e^{i\left( kt+m\phi
\right) }, F_{2}=R_{2}\left( r\right) A_{2}\left( \theta \right) e^{i\left( kt+m\phi
\right) }, \notag
\end{equation}%
\begin{equation}
G_{1}=R_{2}\left( r\right) A_{1}\left( \theta \right) e^{i\left( kt+m\phi
\right) },G_{2}=R_{1}\left( r\right) A_{2}\left( \theta \right) e^{i\left( kt+m\phi
\right) }.  \label{10}
\end{equation}
where $k$ is the frequency and $m$ is the azimuthal quantum number of the wave.
Substituting Eq. (\ref{10}) into Eqs (\ref{D10}), and using separation of
variables, the radial parts of CD equations become 
\begin{equation}
\left( \emph{D}+\frac{1}{r}\right) R_{1}=\frac{l\left( l+1\right) }{r}R_{2},
\label{f11}
\end{equation}%
\begin{equation}
\frac{f}{2}\left( \mathit{D}^{\dag }+\frac{f^{\prime }}{2f}+\frac{1}{r}%
\right) R_{2}=\frac{l\left( l+1\right) }{r} R_{1}.  \label{f12}
\end{equation}%
Redefining the functions $R_{1}$ and $R_{2}$ as follows
\begin{equation}
R_{1}\left( r\right) =\frac{1}{r}P_{1}\left( r\right) ,R_{2}\left( r\right) =%
\frac{\sqrt{f}}{\sqrt{2}r}P_{2}\left( r\right) ,
\end{equation}%
then Eqs. (\ref{f11}, \ref{f12}) transform into,%
\begin{equation}
\left( \frac{d}{dr_{\ast }}+ik\right) P_{1}=l\left( l+1\right) P_{2}\sqrt{%
\frac{f}{r}},  \label{f15}
\end{equation}
\begin{equation}
\left( \frac{d}{dr_{\ast }}-ik\right) P_{2}=l\left( l+1\right) P_{1}\sqrt{%
\frac{f}{r}},  \label{f16}
\end{equation}%
where the tortoise coordinate $r_{\ast }$ is defined as $\frac{d}{dr_{\ast }}%
=f\frac{d}{dr}$.

To this end, assume $U_{+}=P_{1}+P_{2},U_{-}=P_{2}-P_{1}$ then, Eqs. (\ref{f15}, \ref{f16}) transfer to one-dimensional Schr\"{o}dinger like equations 
\begin{equation}
\frac{d^{2}U_{\pm }}{dr_{\ast }^{2}}+\left( k^{2}-V_{\pm }\right) U_{\pm }=0,
\label{s3}
\end{equation}%
where the effective potentials $V_{\pm }$ of massless Dirac field are given
by%
\begin{equation}
V_{\pm }=\frac{l\left( l+1\right) }{r^{2}}\left( l\left( l+1\right) f\pm 
\frac{r\sqrt{f}f^{\prime }}{2}\mp f^{3/2}\right) .  \label{dv1}
\end{equation}

\section{Greybody factors}

The aim of this section is to evaluate the GFs of the regular MOG BH using the general semi-analytic bounds
method. In this method, GFs (or transfer coefficients) of a test field
surrounding a BH should always be greater than or equal to the following
expression: \cite{qn42,qn54,qn55} 
\begin{equation}
\sigma _{l}\left( w\right) \geq \sec h^{2}\left( \frac{1}{2w}%
\int_{r_{h}}^{+\infty }V_{eff}dr_{\ast }\right) ,  \label{is8}
\end{equation}

in which $r_{\ast }$ is the tortoise coordinate. In this process, a
significant role is played by the metric function in determining the
relationship between the GFs and effective potentials. In the previous
section, we obtained potentials for three different perturbations. These
potentials will be used to calculate the GFs. Note, we focus on $V_{+}$ only
in evaluating the GFs.

\subsection{GF of Massless scalar field}

To evaluate the GF of massless scalar field, we  substitute the effective potential (\ref{v1}) into Eq. (\ref{is8}), hence
\begin{equation}
\sigma _{l}\left( w\right) \geq \sec h^{2}\left( \frac{1}{2\omega }%
\int_{r_{h}}^{+\infty }\left( \frac{l(l+1)}{r^{2}}+\frac{2\alpha M^{2}\left(
1+\alpha \right) }{\left( r^{2}+\alpha \left( 1+\alpha \right) M^{2}\right)
^{2}}\right. \right.   \label{in10}
\end{equation}%
\begin{equation*}
\left. \left. -\frac{4M\left( 1+\alpha \right) }{\left( r^{2}+\alpha \left(
1+\alpha \right) M^{2}\right) ^{3/2}}+\frac{6Mr^{2}\left( 1+\alpha \right) }{%
\left( r^{2}+\alpha \left( 1+\alpha \right) M^{2}\right) ^{5/2}}-\frac{%
4\alpha M^{2}r^{2}\left( 1+\alpha \right) }{\left( r^{2}+\alpha \left(
1+\alpha \right) M^{2}\right) ^{3}}\right) dr\right) .
\end{equation*}%
The analytical solution of Eq. (\ref{in10}) is 
\begin{equation*}
\sigma _{l}\left( w\right) \geq \sec h^{2}\left[ \frac{1}{\omega }\left( 
\frac{2l(l+1)}{r_{h}}-\frac{2\alpha M^{2}\left( 1+\alpha \right) r_{h}}{%
\left( r_{h}^{2}+\alpha \left( 1+\alpha \right) M^{2}\right) ^{2}}-\frac{%
r_{h}}{r_{h}^{2}+\alpha \left( 1+\alpha \right) M^{2}}\right) \right. 
\end{equation*}%
\begin{equation}
+\left. \frac{4r_{h}\left( r_{h}^{2}+2\alpha \left( 1+\alpha \right)
M^{2}\right) }{\left( r_{h}^{2}+\alpha \left( 1+\alpha \right) M^{2}\right)
^{3/2}}-\frac{ArcTan\left( \frac{r_{h}}{M\sqrt{\alpha\left(
1+\alpha \right) }}\right) }{M\sqrt{\alpha\left( 1+\alpha \right) }}%
+\frac{2}{M\alpha }-\frac{\pi }{4M\sqrt{\alpha \left( 1+\alpha \right) }}%
\right] .  \label{gfs}
\end{equation}%
Figure 2 depicts the variation of the GF with various parameters $\alpha $.
In this figure, GF is zero for low frequencies, and one for high frequencies, demonstrating that if the frequency is low, the wave can be completely reflected, and at high frequencies, it is not. Moreover, it is seen that as MOG parameter increases, GF increases as well, which allows more thermal radiation to reach the observer at spatial infinity.\begin{figure}
\centering
\includegraphics[scale=1]{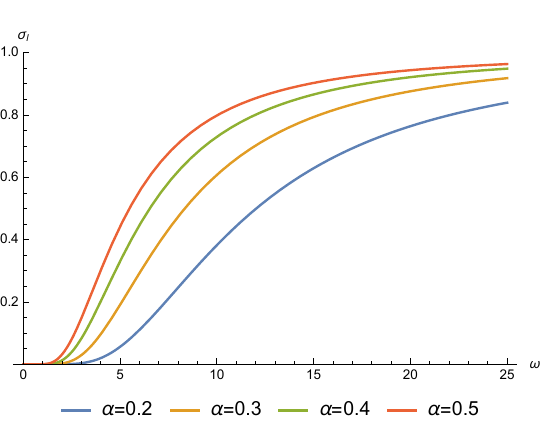} \caption{The greybody bound of scalar massless field of MOG BH for several values of the parameter $\alpha$.}
\label{Figure2}%
\end{figure}

\subsection{GF of gravitational }

To compute the GF of gravitational field, we consider the potential (\ref{vg1}). Thus, Eq. (\ref{is8}) becomes%
\begin{equation}
\sigma _{l}\left( w\right) \geq \sec h^{2}\left( \frac{1}{2\omega }%
\int_{r_{h}}^{+\infty }\left( \frac{l(l+1)}{r^{2}}-\frac{6\alpha M^{2}\left(
1+\alpha \right) }{\left( r^{2}+\alpha \left( 1+\alpha \right) M^{2}\right)
^{2}}\right. \right.   \label{in11}
\end{equation}%
\begin{equation*}
\left. \left. +\frac{12M\left( 1+\alpha \right) }{\left( r^{2}+\alpha \left(
1+\alpha \right) M^{2}\right) ^{3/2}}-\frac{18Mr^{2}\left( 1+\alpha \right) 
}{\left( r^{2}+\alpha \left( 1+\alpha \right) M^{2}\right) ^{5/2}}+\frac{%
12\alpha M^{2}r^{2}\left( 1+\alpha \right) }{\left( r^{2}+\alpha \left(
1+\alpha \right) M^{2}\right) ^{3}}\right) dr\right) .
\end{equation*}%
The analytical solution of Eq. (\ref{in11}) is 
\begin{equation*}
\sigma _{l}\left( w\right) \geq \sec h^{2}\left[ \frac{1}{\omega }\left( 
\frac{2l(l+1)}{r_{h}}+\frac{3\alpha M^{2}\left( 1+\alpha \right) r_{h}}{%
\left( r_{h}^{2}+\alpha \left( 1+\alpha \right) M^{2}\right) ^{2}}+\frac{%
3r_{h}}{r_{h}^{2}+\alpha \left( 1+\alpha \right) M^{2}}\right) \right. 
\end{equation*}%
\begin{equation}
-\left. \frac{6r_{h}\left( r_{h}^{2}+2\alpha \left( 1+\alpha \right)
M^{2}\right) }{\left( r_{h}^{2}+\alpha \left( 1+\alpha \right) M^{2}\right)
^{3/2}}+\frac{3ArcTan\left( \frac{r_{h}}{M\sqrt{\alpha\left(
1+\alpha \right) }}\right) }{M\sqrt{\alpha\left( 1+\alpha \right) }}%
+\frac{6}{M\alpha }-\frac{3\pi }{4M\sqrt{\alpha \left( 1+\alpha \right) }}%
\right] .  \label{gfg}
\end{equation}%
Figure 3 depicts the variation of the greybody bound of gravitational field with different values of MOG parameter $\alpha $.
Figure 3 shows the same behavior as Fig. 2. For larger $\alpha $, the
greybody bound is also larger.\begin{figure}
\centering
\includegraphics[scale=1]{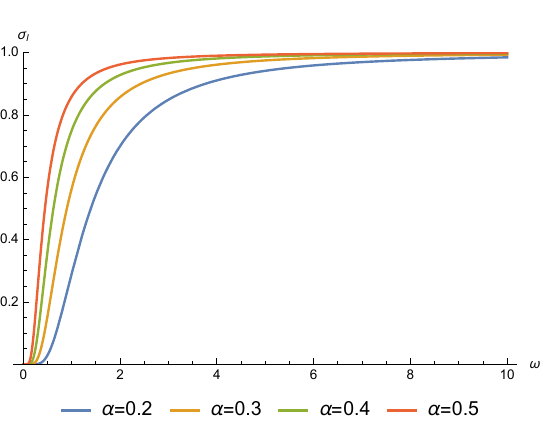} \caption{ The greybody bound of gravitational field of MOG BH for several values of the parameter $\alpha$.}
\label{Figure3}%
\end{figure}

\subsection{GF of massless Dirac }

To compute the GF of massless Dirac field, we consider the potential derived in (\ref{dv1}). Therefore, Eq. (\ref{is8}) becomes
\begin{equation}
\sigma _{l}\left( w\right) \geq \sec h^{2}\left( \frac{1}{2\omega }%
\int_{r_{h}}^{+\infty }\frac{l\left( l+1\right) }{r^{2}}\left( l\left(
l+1\right) + \frac{rf^{\prime }}{2\sqrt{f}}- \sqrt{f}\right) \right) .
\end{equation}%
The complicated nature of the preceding integral precluded an analytical
solution. This problem can be solved by expanding the integrand by an asymptotic series and then evaluating the integral. Thus, the greybody bound of massless Dirac field is 
\begin{equation}
\sigma _{l}\left( w\right) \geq \sec h^{2}\left[ \frac{l\left( l+1\right) }{%
8w}\left( \frac{8\left( l\left( l+1\right) -1\right) }{r_{h}}+\frac{8M\left(
1+\alpha \right) }{r_{h}^{2}}+\frac{4M^{2}\left( 1+\alpha \right) }{r_{h}^{3}%
}\right. \right. .
\end{equation}%
\begin{equation*}
\left. \left. -\frac{4M^{3}\left( 1+\alpha \right) ^{2}\left( 3\alpha
-1\right) }{r_{h}^{4}}-\frac{M^{4}\left( 1+\alpha \right) ^{2}\left( 4\alpha
^{2}+8\alpha -5\right) }{r_{h}^{5}}\right) \right] .
\end{equation*}%
The fluctuation of the GF with different parameters of $\alpha $ is shown in
Fig. 4. The behavior in Fig. 4 is the same as in Fig. 2. The greybody bound
grows along with larger $\alpha $.\begin{figure}
\centering
\includegraphics[scale=1]{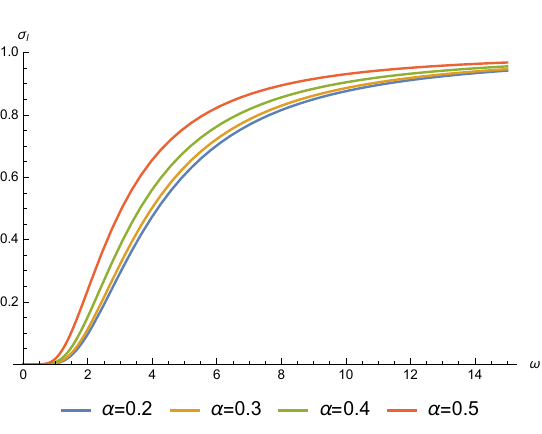} \caption{ The greybody bound of massless Dirac field of MOG BH for several values of the parameter $\alpha$.}
\label{Figure4}%
\end{figure}

\section{QNM}

In studying BH perturbations, QNMs provide a method for understanding BH
behavior under external perturbations, such as scalar, electromagnetic, and
gravitational fields. The QNM is the characteristic frequency at which
gravitational waves are emitted when a BH is perturbed. BH properties
including mass, charge, and spin influence the QNM complex frequencies. Real and imaginary parts of the complex frequency of a QNM
represent the oscillatory frequency of the mode and the decay rate, respectively. Gravitational
waves can provide insight into the physical characteristics of the objects
that created them by observing their QNMs. 
Using the master wave Eqs. (\ref{s1}), (\ref{s2}) and (\ref{s3}) with the boundary conditions of pure outgoing waves at infinity, and pure ingoing waves at the event horizon, we can determine the complex frequency spectrum of the regular MOG BH. Therefore, the solution $U\left(
r_{\ast }\right) $ of the master Eqs. (\ref{s1}), (\ref{s2}) and (\ref%
{s3}) should behave as:%
\begin{equation}
U\sim e^{+iwr_{\ast }},r_{\ast }\rightarrow +\infty ,  \label{co1}
\end{equation}%
\begin{equation*}
U\sim e^{-iwr_{\ast }},r_{\ast }\rightarrow -\infty .
\end{equation*}%
Hence, Quasinormal frequency spectrum is the set of complex frequencies $w$ that
satisfy the master equations and the boundary conditions shown above.

There are several strategies for obtaining QNM frequencies. However, in
this work, we will implement the WKB method of sixth order. The WKB method,
originally proposed by Iyer and Will \cite{qn34} up to third order and later
upgraded to sixth order by Konoplya \cite{qn56}. It is a semianalytic technique for
determining the complex quasinormal mode frequencies of BHs for any type of
field perturbation, including gravitational perturbations. When compared to
other calculation procedures, it already incorporates the boundary
conditions of Eq. (\ref{co1}), and it produces fully  accurate results \cite{qn57}.
The following is the frequency formula for the sixth order WKB method:
\begin{equation}
\omega^{2}=\left[  V_{0}+\sqrt{-2V_{0}^{\prime\prime}}\Lambda\left(  n\right)  -i\beta \sqrt{-2V_{0}^{\prime\prime }}\left(  1+\Omega\left(  n\right)  \right)  \right]  ,\label{isq43}%
\end{equation}
where%
\begin{equation}
\Lambda\left(  n\right)  =\frac{1}{\sqrt{-2V_{0}^{\prime\prime
}}}\left[  \frac{1}{8}\left(  \frac{V_{0}^{\left(  4\right)  }}{V_{0}^{\prime\prime}}\right)  \left(  \frac{1}{4}+\beta^{2}\right)  -\frac{1}{288}\left(
\frac{V_{0}^{\prime\prime\prime}}{V_{0}^{\prime\prime}}\right)  ^{2}\left(  7+60\beta^{2}\right)  \right]  ,\label{isq44}%
\end{equation}
and
\begin{multline}
\Omega\left(  n\right)  =\frac{1}{-2V_{0}^{\prime\prime}}\left[  \frac{5}{6912}\left(  \frac{V_{0}^{\prime\prime\prime}}{V_{0}^{\prime\prime
}}\right)  ^{4}\left(  77+188\beta^{2}\right)  -\frac{1}{384}\left(
\frac{V_{0}^{\prime\prime\prime 2}V_{0}^{\left(  4\right)  }}{V_{0}^{\prime\prime3}}\right)  \left(  51+100\beta^{2}\right)  +\right.  \\
\left.  \frac{1}{2304}\left(  \frac{V_{0}^{\left(  4\right)  }}{V_{0}^{\prime\prime}}\right)  ^{2}\left(  67+68\beta^{2}\right)  +\frac{1}{288}\left(
\frac{V_{0}^{\prime\prime\prime}V_{0}^{\left(  5\right)  }}{V_{0}^{\prime\prime 2}}\right)  \left(  19+28\beta^{2}\right)  -\frac{1}{288}\left(  \frac
{V_{0}^{\left(  6\right)  }}{V_{0}^{\prime\prime
}}\right)  \left(  5+4\beta^{2}\right)  \right], \label{isq45}
\end{multline}
 where $\beta=n+\frac{1}{2}, V_{0}^{(n)}=\frac{d^{n}V(r^0_{\ast})}{dr_{\ast}^{n}}$, and the value of $V_{0}$ indicates the maximum of the effective potential $V_{+}$.
The results of QNM frequencies are listed in Tables \ref{tab1}, \ref{tab2} and \ref{tab3},
respectively. Results show that, all frequencies have a positive real part and a negative imaginary part, indicating that the MOG BH is stable against these perturbations. It is interesting to note that the real and imaginary parts of QNM frequency decrease as the MOG parameter increases, indicating gravitational wave and damping oscillation frequency decreases. 
It is due to the change in potential shapes that the frequencies are altered. We found that the frequency's behavior is comparable to that of quantum mechanics. The faster
the wave decays, the larger the potential as shown in figures 5, 6 and 7. Essentially, a high MoG parameter value smooths out the effective potential (lower peaks) as a result of its large value. The QNM frequency
decreases as the parameter $\alpha $ increases, while the decay rate slows
down. 

\bigskip

\begin{center}
 \begin{tabular}{|c|c|c|}
\hline
$\alpha$ & $n$ & Quasinormal frequency\\
\hline\hline
$0$ & $0$ & $0.966422-0.19361i$ \\ 
& $1$ & $0.926383-0.59162i$ \\ 
$0.1$ & $0$ & $0.907642-0.174542i$ \\ 
& $1$ & $0.874222-0.532551i$ \\ 
$0.2$ & $0$ & $0.857748-0.157992i$ \\ 
& $1$ & $0.829504-0.481319i$ \\ 
$0.3$ & $0$ & $0.814991-0.143304i$ \\ 
& $1$ & $0.790686-0.435877i$ \\ 
$0.4$ & $0$ & $0.7781-0.129948i$ \\ 
& $1$ & $0.756533-0.394587i$ \\ \hline 
\end{tabular}
\captionof{table}{QNM frequencies of MOG BH for massless scalar perturbation.} \label{tab1}
\end{center}
\begin{center}
\begin{tabular}{|c|c|c|}
\hline
$\alpha$ & $n$ & Quasinormal frequency\\
\hline\hline
$0$ & $0$ & $0.746324-0.178435i$ \\ 
& $1$ & $0.692035-0.549831i$ \\ 
$0.1$ & $0$ & $0.707269-0.16037i$ \\ 
& $1$ & $0.66344-0.493112i$ \\ 
$0.2$ & $0$ & $0.674727-0.144872i$ \\ 
& $1$ & $0.639892-0.444491i$ \\ 
$0.3$ & $0$ & $0.647504-0.13122i$ \\ 
& $1$ & $0.620247-0.40176i$ \\ 
$0.4$ & $0$ & $0.624772-0.118777i$ \\ 
& $1$ & $0.603509-0.362927i$\\ \hline
\end{tabular}
\captionof{table}{QNM frequencies of MOG BH for gravitational perturbation.} \label{tab2}
\end{center}
\begin{center}  
\begin{tabular}{|c|c|c|}
\hline
$\alpha$ & $n$ & Quasinormal frequency\\
\hline\hline
$0$ & $0$ & $0.757254-0.193085i$ \\ 
& $1$ & $0.707208-0.597493i$ \\ 
$0.1$ & $0$ & $0.712145-0.174078i$ \\ 
& $1$ & $0.67105-0.536864i$ \\ 
$0.2$ & $0$ & $0.673769-0.15755i$ \\ 
& $1$ & $0.639487-0.484386i$ \\ 
$0.3$ & $0$ & $0.640813-0.142852i$ \\ 
& $1$ & $0.611555-0.437912i$ \\ 
$0.4$ & $0$ & $0.612312-0.129456i$ \\ 
& $1$ & $0.586391-0.395726i$\\ \hline
 \end{tabular}
 \end{center}
\captionof{table}{QNM frequencies of MOG BH for massless Dirac perturbation.} \label{tab3}

\begin{figure}
    \centering
{{\includegraphics[width=7.5cm]{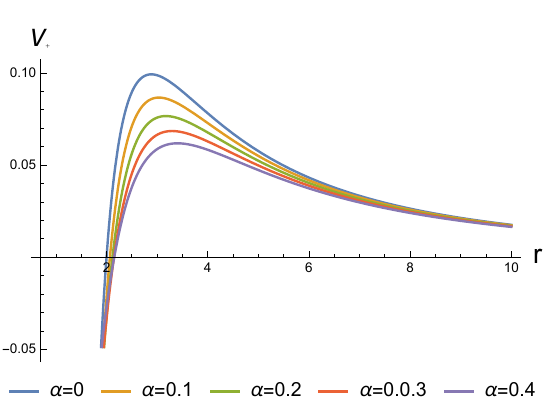} }}\qquad
    {{\includegraphics[width=7.5cm]{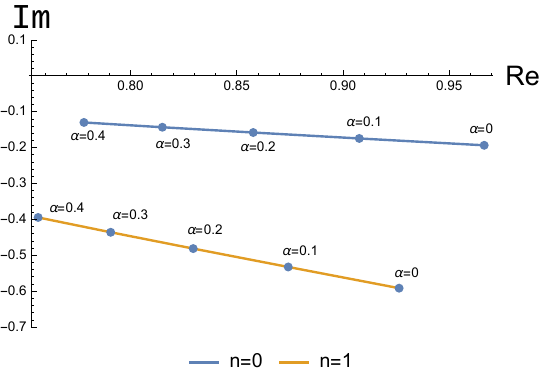}}}
    \caption{Effective potentials (\ref{v1})
for massless scalar field for different values of MOG parameter $\alpha$ (left). The corresponding QNM (right).
Here, $l =  1 = M  $.}%
    \label{fig5}%
\end{figure}
\begin{figure}
    \centering
{{\includegraphics[width=7.5cm]{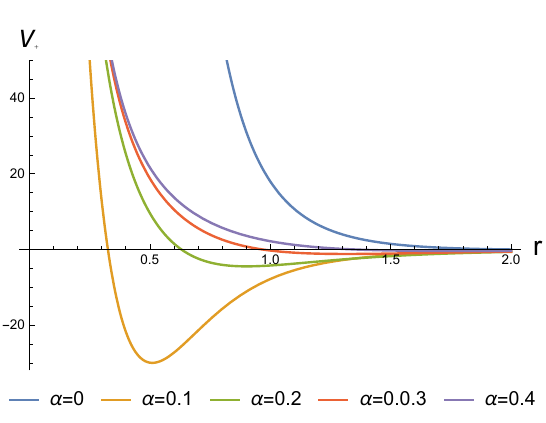} }}\qquad
    {{\includegraphics[width=7.5cm]{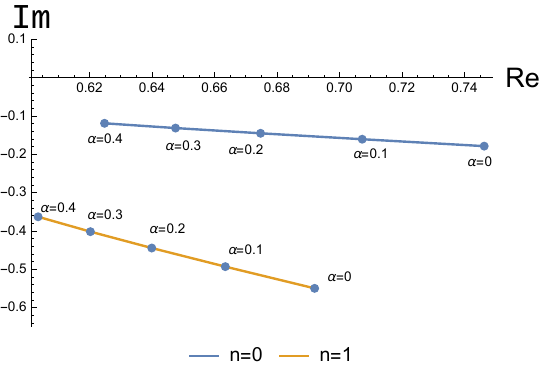}}}
    \caption{Effective potentials (\ref{vg1})
for axial gravitational field for different values of MOG parameter $\alpha$ (left). The corresponding QNM (right).
Here, $l =  2 = M  $.}%
    \label{fig6}%
\end{figure}

 \begin{figure}
    \centering
{{\includegraphics[width=7.5cm]{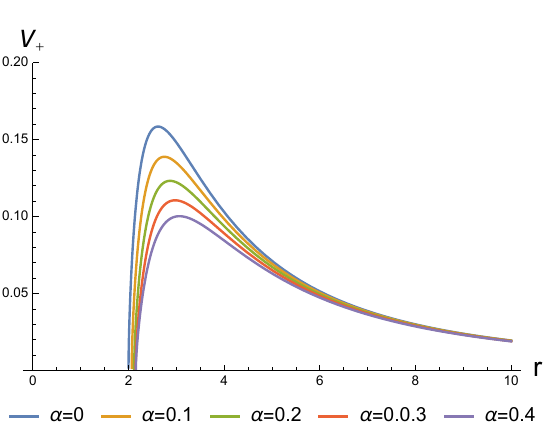} }}\qquad
    {{\includegraphics[width=7.5cm]{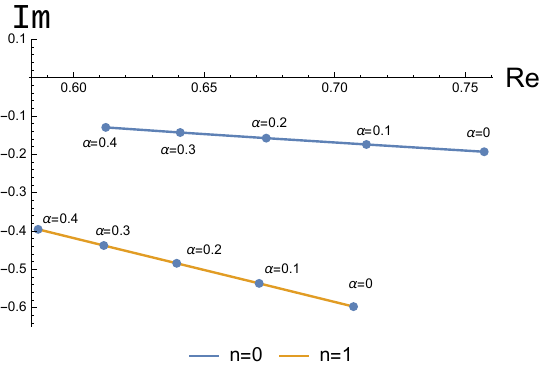}}}
    \caption{Effective potentials (\ref{dv1})
for massless Dirac field for different values of MOG parameter $\alpha$ (left). The corresponding QNM (right).
Here, $l =  1 = M  $.}%
    \label{fig7}%
\end{figure}

\section{Conclusion}

In this paper, we studied the GF and QNM of the regular MOG BH. Studying the massless scalar field, axial 
gravitational, and massless Dirac perturbations of MOG BH helps us understand BH
behavior in MOG theory and differentiate it from GR. For three perturbations of 
BH, we derived the effective potential equations. The bounds of GFs were calculated using the derived potentials. As a next step, we applied the WKB method of sixth order to the MOG BH QNM. 
Additionally, we have analyzed the influence of the MOG parameter graphically. 
All the results were shown as a function of the MOG parameter of the BH. All the plots
were made by varying the values of the MOG parameter in a step of 0.1 in $\alpha$. 
These are the main conclusions of this paper.

$\bullet $ We have found that as the parameter $\alpha$ increases the greybody bounds increase as well (Figures 2, 3 and 4). Therefore, higher MOG parameter values of MOG BH will result in a higher probability of detecting Hawking radiation.

$\bullet$ It has been demonstrated that MOG parameter $\alpha$ and eigenvalue $l$ can dramatically influence the shape of the effective potential.

$\bullet$ It is found that increasing the value of parameter $\alpha$ reduces the relative change of the effective potential, resulting in lower peak values and a decrease in gravitational wave oscillation frequency and damping.

$\bullet $ Our results showed that all frequencies have a positive real part and a negative imaginary part, demonstrating that the MOG BH is stable against these perturbations. These frequencies correspond to the ringdown profile of BH mergers and the damping rate of propagating gravitational waves at the end of the waveform. 

$\bullet $ As the parameter increases, the real and imaginary frequencies of the QNM frequency decrease. This implies that the gravitational wave oscillation frequency and damping decrease as well.

$\bullet $ We found that the behavior of frequencies is comparable to quantum mechanics. The larger the potential, as shown in figures 5, 6, and 7, the faster the wave decays.

In this work, perturbations in a static spherically symmetric MOG spacetime were studied. The work can be extended to a regular rotating MOG dark compact object in the future \cite{qn25}. In addition, it is interesting to explore the influence of MOG parameter and angular momentum (spin) on GF and QNM for the BH we study. 

\bigskip
{\Large Data Availability}
\newline No data availability in this manuscript.

\end{document}